# Robust 3.7 V-$Na_{2/3}[Cu_{1/3}Mn_{2/3}]O_2$ Cathode for Na-ion Batteries

*Xiaohui Rong[1,2,3], Xingguo Qi[1,4], Quan Zhou[1,4], Libin Kang[4], Dongdong Xiao[1,5], Ruijuan Xiao[1,2], Feixiang Ding[1], Yang Yang[1], Yuan Liu[1], Yun Su[1], Shiguang Zhang[1], Lunhua He[1,6,7], Yaxiang Lu[1,2,5], Liquan Chen[1,2,3,5], Yong-Sheng Hu[1,2,3,4,5],\**

[1] Key Laboratory for Renewable Energy, Beijing Key Laboratory for New Energy Materials and Devices, Beijing National Laboratory for Condensed Matter Physics, Institute of Physics, Chinese Academy of Sciences, Beijing, China.

[2] College of Materials Science and Optoelectronic Technology, University of Chinese Academy of Sciences, Beijing 100049, China.

[3] Yangtze River Delta Physics Research Center Co. Ltd, Liyang 213300, China.

[4] HiNa Battery Technology Co, Ltd, Beijing, China

[5] Huairou Division, Institute of Physics, Chinese Academy of Sciences, Beijing 101400, China.

[6] Songshan Lake Materials Laboratory, Dongguan, China.

[7] Spallation Neutron Source Science Center, Dongguan 523808, China

*Corresponding author: Yong-Sheng Hu (yshu@iphy.ac.cn)

**Abstract**

Na-ion batteries (NIBs), which are recognized as a next-generation alternative technology for energy storage, still suffer from commercialization constraints due to the lack of low-cost, high-performance cathode materials. Since our first discovery of $Cu^{3+}/Cu^{2+}$ electrochemistry in 2014, numerous Cu-substituted/doped materials have been designed for NIBs. However for almost ten years, the potential of $Cu^{3+}/Cu^{2+}$ electrochemistry has been grossly underappreciated and normally regarded as a semielectrochemically active redox. Here, we re-synthesized P2-$Na_{2/3}[Cu_{1/3}Mn_{2/3}]O_2$ and reinterpreted it as a high-voltage, cost-efficient, air-stable, long-life, and high-rate cathode material for NIBs, which demonstrates a high operating voltage of 3.7 V and a completely active $Cu^{3+}/Cu^{2+}$ redox reaction. The 2.3 Ah cylindrical cells exhibit excellent cycling (93.1% capacity after 2000 cycles), high rate (97.2% capacity at 10C rate), good low-temperature performance (86.6% capacity at -30°C), and high safety, based on which, a 56 V-11.5 Ah battery pack for E-bikes is successfully constructed, exhibiting stable cycling (96.5% capacity at the 800$^{th}$ cycle) and a long driving distance (36 km, tester weight 65 kg). This work offers a commercially feasible cathode material for low-cost, high-voltage NIBs, paving the way for advanced NIBs in power and stationary energy storage applications.

**Introduction**

Na-ion batteries (NIBs) are currently one of the most promising electrochemical energy storage technologies. The deintercalation/intercalation behavior of Na ions in $Na_xCoO_2$||Na cells was discovered for the first time in 1980[1]. In the same year, the deintercalation/intercalation behavior of Li ions in $Li_xCoO_2$||Li cells was also reported[2], which showed a much higher voltage (~4.0 V) than that of the $Na_xCoO_2$||Na cells (~3.0 V). The electrochemical advantages of LIBs have motivated investigators to search for other feasible materials and their implementation for industrialization. The commercialization of Li-ion batteries (LIBs) in 1991 further led to the flourishing of LIB research while the research of NIB was a bit lagged. Since the 2010s, with the increasing concern of lithium resource shortages and the demands of grid-scale stationary energy storage, a new era of NIBs has been approaching.

It has been generally established that layered oxide cathode materials with high volumetric energy density and good processability are viable for industrial applications. The development of layered oxide cathode materials has been hindered by some of their limitations, such as: 1) low stability in humid air, which presents in most O3-type layered oxides and needs further modification; 2) low operating voltage (more than 0.3 V lower than that of LIB layered oxides), which restricts direct use of NIBs as an alternative to LIBs and limits the energy density improvements; and 3) wide voltage span, which limits the practical applications (e.g., power batteries in E-bikes). Therefore, it is vital to create substitute cathode materials free of these drawbacks.

The first Cu-based cathode material for NIBs is P2-$Na_{2/3}[Cu_{1/3}Mn_{2/3}]O_2$[3], reported by our group in 2014, showing a reversible capacity of ~70 mAh g$^{-1}$ in the voltage range of 2.5-4.2 V (~50 mAh g$^{-1}$ in 3.0-4.2 V). After that, many Cu-based cathode materials were reported, including P2-$Na_{7/9}[Cu_{2/9}Fe_{1/9}Mn_{2/3}]O_2$[4], O3-$Na_{0.9}[Cu_{0.22}Fe_{0.30}Mn_{0.48}]O_2$[5], and O3-$Na[Cu_{1/9}Ni_{2/9}Fe_{1/3}Mn_{1/3}]O_2$[6], which showed higher reversible specific capacity than P2-$Na_{2/3}[Cu_{1/3}Mn_{2/3}]O_2$. Since then, Cu-based layered oxides have become one of the most important classes of NIB cathode materials. However, the real electrochemical performance of P2-$Na_{2/3}[Cu_{1/3}Mn_{2/3}]O_2$ has not been reported over a nine-year period.

In this work, we identify P2-Na$_{2/3}$[Cu$_{1/3}$Mn$_{2/3}$]O$_2$ as a high voltage (average operating voltage is ~3.7 V) cathode material with a reversible specific capacity of ~83 mAh g$^{-1}$ (2.5-4.1 V). Studies on the synthesis conditions are meticulously carried out, confirming the importance of homogeneity degree of precursors, calcination temperature, and cooling rate for P2-Na$_{2/3}$[Cu$_{1/3}$Mn$_{2/3}$]O$_2$. The narrow voltage range, long cycling life, and good rate performances of Ah-level full cells motivated us to manufacture a 56 V-11.5 Ah battery pack, which also exhibts a long cycling life (96.5% at the 800$^{th}$ cycle), demonstrating the superiority of P2-Na$_{2/3}$[Cu$_{1/3}$Mn$_{2/3}$]O$_2$ for advanced NIBs.

**Results and discussion**

**Structure of P2-Na$_{2/3}$[Cu$_{1/3}$Mn$_{2/3}$]O$_2$**

The P2-Na$_{2/3}$[Cu$_{1/3}$Mn$_{2/3}$]O$_2$ samples were prepared by solid-state reactions using Na$_2$CO$_3$, CuO, and MnO$_2$ as precursors. Two samples with different synthesis conditions were denoted as CM-M900F and CM-B800S, respectively. The neutron diffraction (ND) pattern of the P2-Na$_{2/3}$[Cu$_{1/3}$Mn$_{2/3}$]O$_2$ (CM-B800S) powder sample shown in Fig. 1a could be well refined with the P2 structure with space group $P6_322$, which has honeycomb cation ordering of Cu and Mn on the √3a × √3a superlattice. In contrast, the X-ray diffraction (XRD) pattern of P2-Na$_{2/3}$[Cu$_{1/3}$Mn$_{2/3}$]O$_2$ (CM-B800S) could not easily identify the space group, which could be well refined with either $P6_322$ or $P6_3/mmc$ space groups, different from the cases of P2-Na$_x$[Li,Mn]O$_2$ and P2-Na$_x$[Mg,Mn]O$_2$ compounds[7,8]. The Rietveld refinement result gives lattice parameters of $a = b = 5.015$ Å, $c = 11.16$ Å, and $V = 243.0$ Å$^3$, and the theoretical density is calculated to be 4.388 g cm$^{-3}$. The local Cu-Mn(2)-Cu ordering in the transition metal (TM) layers along the [120] orientation is also confirmed by high-angle annular dark-field-scanning transmission electron microscopy (HAADF-STEM) images (Fig. 1b), in which Cu with a higher atomic number (29) shows higher Z-contrast than Mn (25). Figure 1c illustrates the crystal structure of P2-Na$_{2/3}$[Cu$_{1/3}$Mn$_{2/3}$]O$_2$ ($P6_322$), showing intralayer Cu/Mn honeycomb ordering and interlayer Cu-Cu zig-zag sequence. The sodium ions in the prismatic sites promise fast ion diffusion kinetics, which has been widely demonstrated[9,10].

**Electrochemical and physical properties of P2-Na$_{2/3}$[Cu$_{1/3}$Mn$_{2/3}$]O$_2$**

The charge/discharge voltage profiles of two selected samples, CM-M900F and CM-B800S, are compared in Fig. 1d, showing distinct features. The detailed synthesis conditions of the two materials are introduced in the figure caption of Fig. 1 and the Methods section. The charge/discharge profile of CM-M900F is similar to those of previously reported works[3], with a long slope delivering only ~50 mAh g$^{-1}$ above 3.0 V, according to which some researchers regard Cu as a semielectrochemically active metal[11].

In contrast, the typical charge/discharge profile of the CM-B800S sample shows a high average voltage (~3.7 V vs. Na$^+$/Na, one of the highest values in layered oxide cathodes) and a discharge capacity of ~83 mAh g$^{-1}$ (theoretically 85 mAh g$^{-1}$ considering the Cu$^{3+}$/Cu$^{2+}$ redox reaction). The charge/discharge behaviors (1$^{st}$ cycle, 1.5-4.1 V) of P2-Na$_{2/3}$[Cu$_{1/3}$Mn$_{2/3}$]O$_2$ (CM-B800S) and P2-Na$_{2/3}$[Ni$_{1/3}$Mn$_{2/3}$]O$_2$ are compared in Fig. 1e. The average voltage (2.5-4.1 V) of P2-Na$_{2/3}$[Cu$_{1/3}$Mn$_{2/3}$]O$_2$ (CM-B800S) is much higher (~0.3 V) than that of P2-Na$_{2/3}$[Ni$_{1/3}$Mn$_{2/3}$]O$_2$, while the similar profile shape indicates the same structural transition during Na$^+$ deintercalation/intercalation. When discharging to 1.5 V, only ~25 mAh g$^{-1}$ (~0.09 Na$^+$ intercalation) is delivered by P2-Na$_{2/3}$[Cu$_{1/3}$Mn$_{2/3}$]O$_2$ (CM-B800S), much lower than that of P2-Na$_{2/3}$[Ni$_{1/3}$Mn$_{2/3}$]O$_2$ (~25 mAh g$^{-1}$, ~0.32 Na$^+$ intercalation). This could be attributed to the Jahn-Taller active nature of Cu$^{2+}$, which could restrain the transformation of Jahn-Taller inactive Mn$^{4+}$ to Jahn-Taller active Mn$^{3+}$.

Based on the electrochemistry results, the cost-effectiveness of P2-Na$_{2/3}$[Cu$_{1/3}$Mn$_{2/3}$]O$_2$ (CM-B800S) has thus been evaluated. The energy costs per kWh of three typical lithium battery cathodes (LiFePO$_4$, LiMn$_2$O$_4$, and Li[Ni,Co,Mn]O$_2$) and P2-Na$_{2/3}$[Cu$_{1/3}$Mn$_{2/3}$]O$_2$ (CM-B800S) are compared in Fig. 1f, in which P2-Na$_{2/3}$[Cu$_{1/3}$Mn$_{2/3}$]O$_2$ (CM-B800S) shows a significant advantage (~12 $ kWh$^{-1}$). The high working voltage, superior rate capability, and outstanding cost-effectiveness demonstrated Na$_{2/3}$[Cu$_{1/3}$Mn$_{2/3}$]O$_2$ as an alternative material for high-power batteries.

**Structural evolution upon Na$^+$ deintercalation/intercalation**

*In situ* electrochemical XRD and *ex situ* HAADF-STEM tests are carried out to understand the long-range and short-range structural evolution during Na$^+$ deintercalation/intercalation. A Swaglock cell with P2-Na$_{2/3}$[Cu$_{1/3}$Mn$_{2/3}$]O$_2$ (CM-B800S) on the Al window and Na metal as the anode is charged and discharged at 10 mA g$^{-1}$ in the voltage range of 2.5-4.1 V for collecting XRD profiles. Figure 2a shows the contour plot of the *in situ* electrochemical XRD during the first cycle. During charging to 4.1 V, the (002) and (004) peaks significantly shift to a lower degree, revealing the expansion of the *c*-axis during Na$^+$ deintercalation. Notably and unexpectedly, the two two-phase transitions are confirmed by the step-like shift of the (004), (110), (111), (104), and (113) peaks during both charge and discharge, with each transition corresponding to an inflection point in the voltage profile. The XRD profile of the charged state (4.1 V) could be identified as a P2-type layered oxide with a space group of *P*6$_3$/*mcm* different from that of the pristine state (*P*6$_3$22), according to the formation of (112) and (006) peaks. The two two-phase transitions can also be demonstrated by the calculated lattice parameters (*a* axis, *c* axis, and unit cell volume) shown in Figs. 2b-c. A significant change could be observed in all three parameters, and the phase evolutions during charge and discharge seem asymmetric, which could hardly be explained by the solution-solution reaction and the Na$^+$/vacancy ordering transformation[12]. Although phase transitions occur during the first charge/discharge, variations in the *a* axis (2.07%), *c* axis (2.20%), and unit cell volume (2.78%) are limited, which promises the steady cycling of P2-Na$_{2/3}$[Cu$_{1/3}$Mn$_{2/3}$]O$_2$ (CM-B800S).

The *ex situ* HAADF-STEM images provide direct evidence and atom-level details of the phase transition. As demonstrated above, due to the atomic number difference between Cu and Mn, intralayer Cu/Mn honeycomb ordering and interlayer Cu-Cu zig-zag sequences are found in the pristine sample (Fig. 2e). The intralayer Cu/Mn honeycomb is kept during the whole charge process, while the interlayer Cu-Cu sequence changes twice. For the sample first charged to 3.8 V, both Cu-Cu zig-zag and linear sequences could be found in the HAADF-STEM image, which could be a coexisting phase with one of the *P6₃22* space groups and another of the *P6₃/mcm* space group. This new phase, which we name the PP4 phase in this study, differs from the OP4 phase or OP2 phase, in which Na+ ions occupy both octahedral and prismatic sites. When charged to 4.1 V, the interlayer Cu-Cu sequence transforms to a linear sequence, which is classified as the *P63/mcm* space group and consistent with the *in situ* XRD result.

**Behaviors of P2-Na$_{2/3}$[Cu$_{1/3}$Mn$_{2/3}$]O$_2$ in Ah-level cells and packs**

Ah-level cell/battery performance is identified as the "golden standard" to evaluate the material's potential for practical applications. For batch fabrication of 26700 cylindrical cells, kilograms of P2-Na$_{2/3}$[Cu$_{1/3}$Mn$_{2/3}$]O$_2$ (CM-B800S) was synthesized via the same condition as that used for the gram-level sample. More than 200 cylindrical cells (26700 type) with P2-Na$_{2/3}$[Cu$_{1/3}$Mn$_{2/3}$]O$_2$ as the cathode material and hard carbon as the anode material were successfully fabricated, delivering ~2.3 Ah (~90 Wh kg$^{-1}$) reversible capacity at 0.2C/0.2C with a weight of ~90.5 g. Figure 3a shows the charge/discharge voltage profiles of a 26700 cylindrical cell tested at various rates in the voltage range of 2.0-4.05 V, showing a high capacity retention of 97.2% and a high energy retention of 82.9% at a 10C rate (0.2C as 100%). The capacity retention, energy retention, and median discharge voltage of the rate performance are summarized in Table 1. The steady energy output is another important requirement for practical application. The cell could stably work at a high temperature of 60°C, and a high capacity retention of 86.6% and a high energy retention of 73.4% at -30°C (25°C as 100%) were confirmed, promising great adaptability to different circumstances (Fig. 3b, Table 1). The 26700 cylindrical cell also shows long-term cycling stability, as shown in Figs. 3c and 3d, showing a high capacity retention of 93.1% after 2000 cycles (2C/2C, 2.0-4.05 V, 25°C). More importantly, 26700 cylindrical cells passed various rigorous safety tests, including external shorting, overcharging (1C, 6 V), and high-temperature storage (130°C, 30 min), showing excellent safety performance.

Based on the performance characteristics of the 26700 cylindrical cells, a 56 V-11.5 Ah battery pack was successfully fabricated, which was applied to two-wheeled

E-bikes. The battery pack shows stable cycling for more than 800 charge/discharge cycles (96.5% capacity retention at the 800$^{th}$ cycle), as shown in Figs. 5e-g. The C-rate is set as 0.5C/0.5C (30-60.75 V, 25°C), which is close to that required by practical usage. Actual on-road testing is carried out using an E-bike equipped with a 350 W motor and driven by a tester weighing 65 kg. An endurance mileage of ~36 km, close to the commercialized 48 V-12 Ah Li-ion battery pack, demonstrates great potential for next-generation batteries.

In conclusion, a 3.7 V high voltage Cu-Mn-based layered cathode is developed for advanced Na-ion batteries, which shows a similar voltage range as LIBs and could be perfectly employed in the devices driven by LIBs. The ND, *in situ* electrochemistry XRD, and *ex situ* HAADF-STEM results demonstrate the transformations of interlayer honeycomb stacking sequences at ~3.7 V and ~4.0 V, corresponding to the two-voltage plateau character in 2.5-4.1 V (vs. Na$^+$/Na). Mass-manufactured 26700 cylindrical full cells (P2-Na$_{2/3}$[Cu$_{1/3}$Mn$_{2/3}$]O$_2$||HC) exhibit superior cycling, rate, safety, and low-temperature discharge performances. According to the multidimensional performance assessment, P2-Na$_{2/3}$[Cu$_{1/3}$Mn$_{2/3}$]O$_2$ has even superior comprehensive performance than widely used O3-Na[Ni$_{1/3}$Fe$_{1/3}$Mn$_{2/3}$]O$_2$ (Fig. 3h). The successful application of the long-life 56 V-11.5 Ah battery pack in a two-wheel E-bike further demonstrates the great potential of P2-Na$_{2/3}$[Cu$_{1/3}$Mn$_{2/3}$]O$_2$ for power and stationary energy storage battery applications, which will undoubtedly boost the confidence of all NIB investigation in academia and industry worldwide

**Methods**

**Materials synthesis**

All the samples were prepared by a solid-state reaction using a precursor of $Na_2CO_3$ (99.9%, Alfa), CuO (98%, Alfa), and $MnO_2$ (99.9%, Alfa). The starting materials were weighed and ground in an agate mortar according to the appropriate stoichiometric ratio (2% more $Na_2CO_3$ was used because of volatilization loss during the process). Six samples with different conditions were synthesized. The sample CM-B800S was synthesized with a ball-milled precursor and was calcined at 800°C before slow cooling in air. The sample CM-M900F was synthesized with a manually milled precursor and was calcined at 900°C before fast cooling in air.

**Cell and battery tests**

All electrochemical tests were conducted using coin cells (CR2032) assembled in an argon-filled glovebox. An 80 wt% active material, 10 wt% Super P, and 10 wt% polyvinylidene fluoride (PVdF) were used to prepare working electrodes with Al foil as the current collector. The loading mass of active material on the positive electrode was controlled to between 3.0 and 4.0 mg/cm$^{-2}$. The electrolyte was a solution of 1 M $NaClO_4$ in ethylene carbonate (EC):propylene carbonate (PC):dimethyl carbonate (DMC) (1:1:1 in volume) and fluoroethylene carbonate (2% in volume). The 1 M $NaPF_6$ in ethylene carbonate (EC): diethyl carbonate (DEC) with additives electrolyte was used in the 26700 cylindrical full cells. Sodium foil was used as the counter electrode, and glass fiber was used as the separator for the coin half cells. The pyrolyzed anthracite anode (HiNa Co., Ltd.) was used for the 26700 cylindrical full-cell test. The charge and discharge tests of coin cells, 26700 cylindrical cells, and battery packs were carried out using Neware battery test systems at set temperatures.

**Material characterizations**

*X-ray diffraction and neutron diffraction*

The structure was characterized using an X-ray diffractometer (D8 Bruker) with Cu-Kα radiation (λ = 1.5405 Å) in the scan range (2θ) of 10° to 80°. The *in situ* heating *X-ray diffraction* (XRD) measurements were performed from room temperature to 900°C, with the heating rate set to 5°C min$^{-1}$ and a holding time of 10 min for each temperature point. A specially designed Swagelok cell equipped with an X-ray-transparent aluminum window was used for the *in situ* electrochemical XRD studies. The *in situ* XRD patterns were collected with an interval of 30 min for each scan from 10° to 75° on charge and discharge at a current density of 10 mg cm$^{-2}$, between 2.5-4.1 V and 1.5-4.6 V versus Na/Na$^+$. The neutron data were measured at the General Purpose Powder Diffractometer (GPPD) at the China Spallation Neutron Source. GSAS2 was used to refine the XRD and neutron data.

*Scanning transmission electron microscopy and scanning electron microscopy*

A JEM-ARM200F scanning transmission electron microscope fitted with a double aberration corrector for both probe-forming and imaging lenses was used to perform high-angle annular dark-field (HAADF) imaging, operated at 200 kV. The convergence angle was 25 mrad, and the angular range of collected electrons for HAADF imaging was approximately 70-250 mrad. The morphologies of the samples were investigated by a scanning electron microscope (Hitachi-S4800).

**Acknowledgments**

This work was supported by the National Natural Science Foundation (NSFC) of China (52002394 and 52122214) and the Young Elite Scientists Sponsorship Program by CAST (2022QNRC001). The authors wish to thank facility support of the 4B9A beamline of Beijing Synchrotron Radiation Facility (BSRF) and General Purpose Powder Diffractometer (GPPD) at the China Spallation Neutron Source.

**Competing interests**

The authors declare no competing interests.

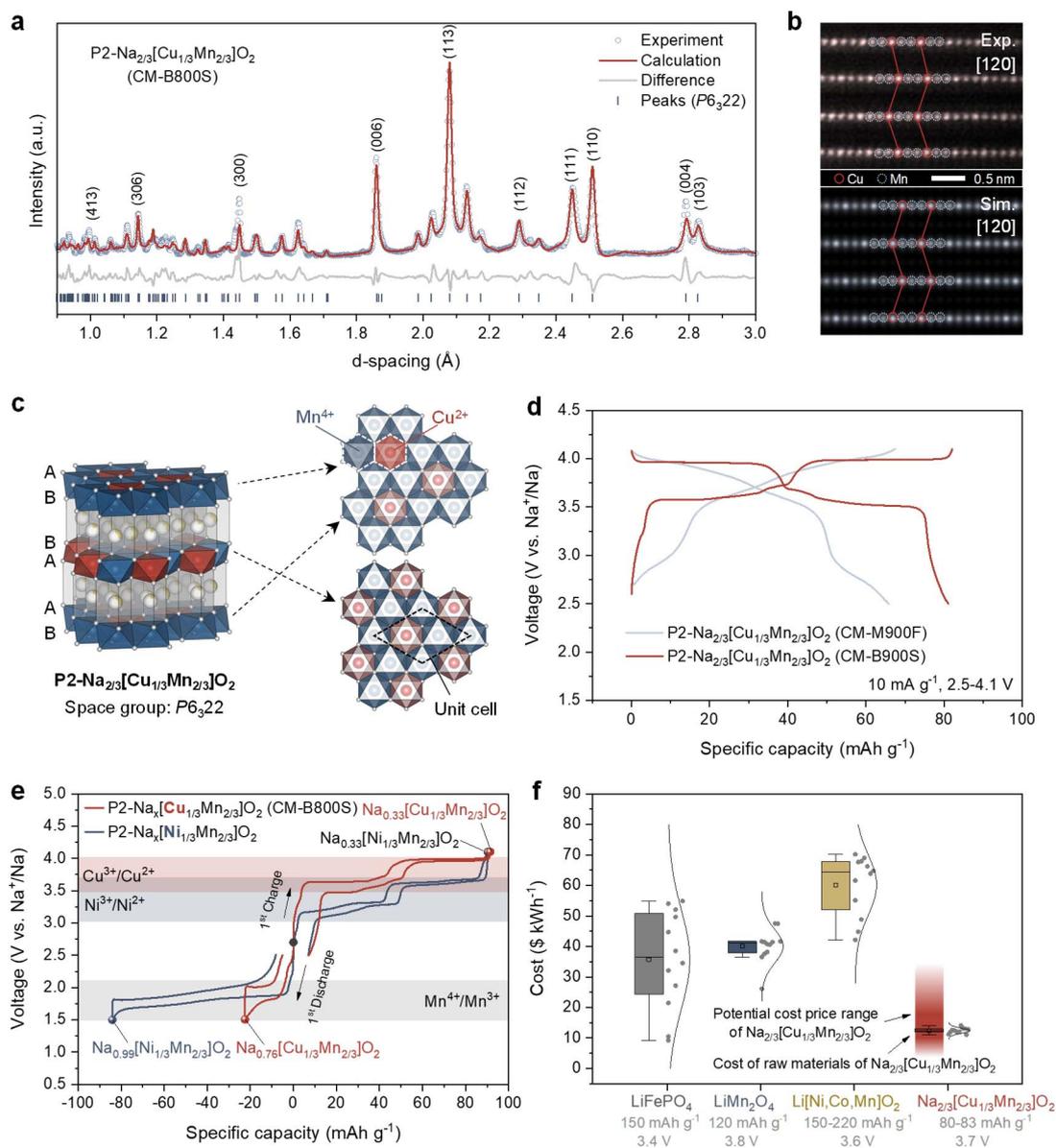

**Fig. 1 | Structure, electrochemistry, and cost of P2-Na$_{2/3}$[Cu$_{1/3}$Mn$_{2/3}$]O$_2$. a**, Refinement result of the neutron diffraction data of P2-Na$_{2/3}$[Cu$_{1/3}$Mn$_{2/3}$]O$_2$ (CM-B800S), whose precursor was ball-milled and calcined at 800°C prior to slow cooling. **b**, Experimentally collected and simulated high-angle annular dark-field-scanning transmission electron microscopy (HAADF-STEM) images of CM-B800S indicate honeycomb ordering in the transition metal layers (····-Cu-2Mn-Cu-····). **c**, Illustration of the ideal P2-Na$_{2/3}$[Cu$_{1/3}$Mn$_{2/3}$]O$_2$ structure. **d**, Significantly different voltage profiles (2$^{nd}$ cycle) of two P2-Na$_{2/3}$[Cu$_{1/3}$Mn$_{2/3}$]O$_2$ samples synthesized under different conditions. The precursor for CM-M900F was milled manually and calcined at 900°C, followed by fast (natural) cooling. **e**, Comparison of the initial voltage profiles between P2-Na$_{2/3}$[Cu$_{1/3}$Mn$_{2/3}$]O$_2$ and P2-Na$_{2/3}$[Cu$_{1/3}$Mn$_{2/3}$]O$_2$ in the voltage range of 1.5-4.1 V. **f**, Comparison of the energy costs per kWh between three typical lithium battery cathodes (LiFePO$_4$, LiMn$_2$O$_4$, and Li[Ni,Co,Mn]O$_2$) and P2-Na$_{2/3}$[Cu$_{1/3}$Mn$_{2/3}$]O$_2$ (publicly available data from the official website of General Administration of Customs of the Peoples Republic of China and chemicalbook.com, Jan. 2022 to Dec. 2022).

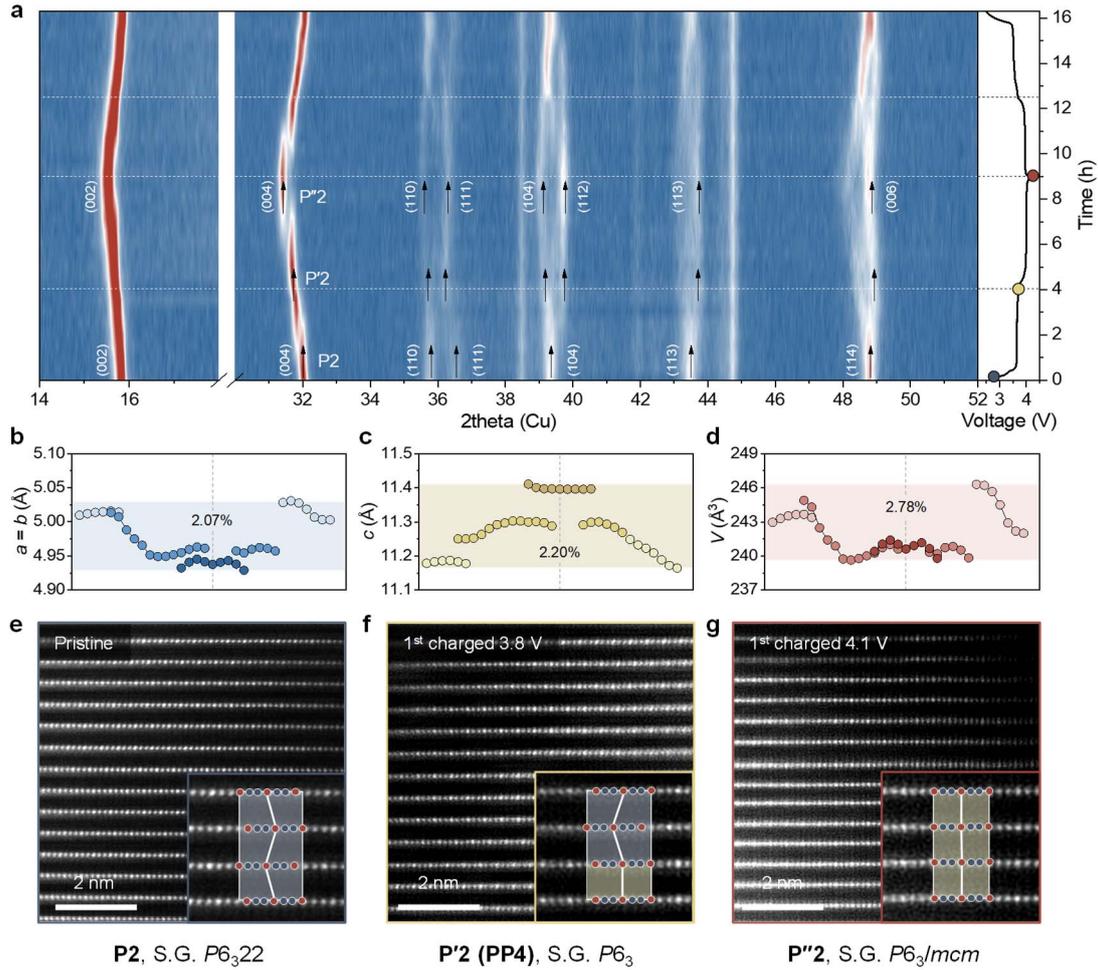

**Fig. 2 | Structure evolution during P2-Na$_{2/3}$[Cu$_{1/3}$Mn$_{2/3}$]O$_2$ during Na$^+$ intercalation and deintercalation. a-d**, *In situ* electrochemical XRD data during the first cycle for P2-Na$_{2/3}$[Cu$_{1/3}$Mn$_{2/3}$]O$_2$. *In situ* electrochemical XRD contour plot (**a**) and evolution of the lattice parameters during the charge/discharge process (*a* axis (**b**), *c* axis (**c**), and unit cell volume (**d**)) indicating two two-phase transition mechanism. **e-g**, *ex situ* HAADF-STEM images of various charge states: pristine (**e**), 1$^{st}$ charged 3.8 V (**f**), and 1$^{st}$ charged 4.1 V (**g**), revealing the topological phase transition origin of the two two-phase transitions found in the *in situ* electrochemical XRD test.

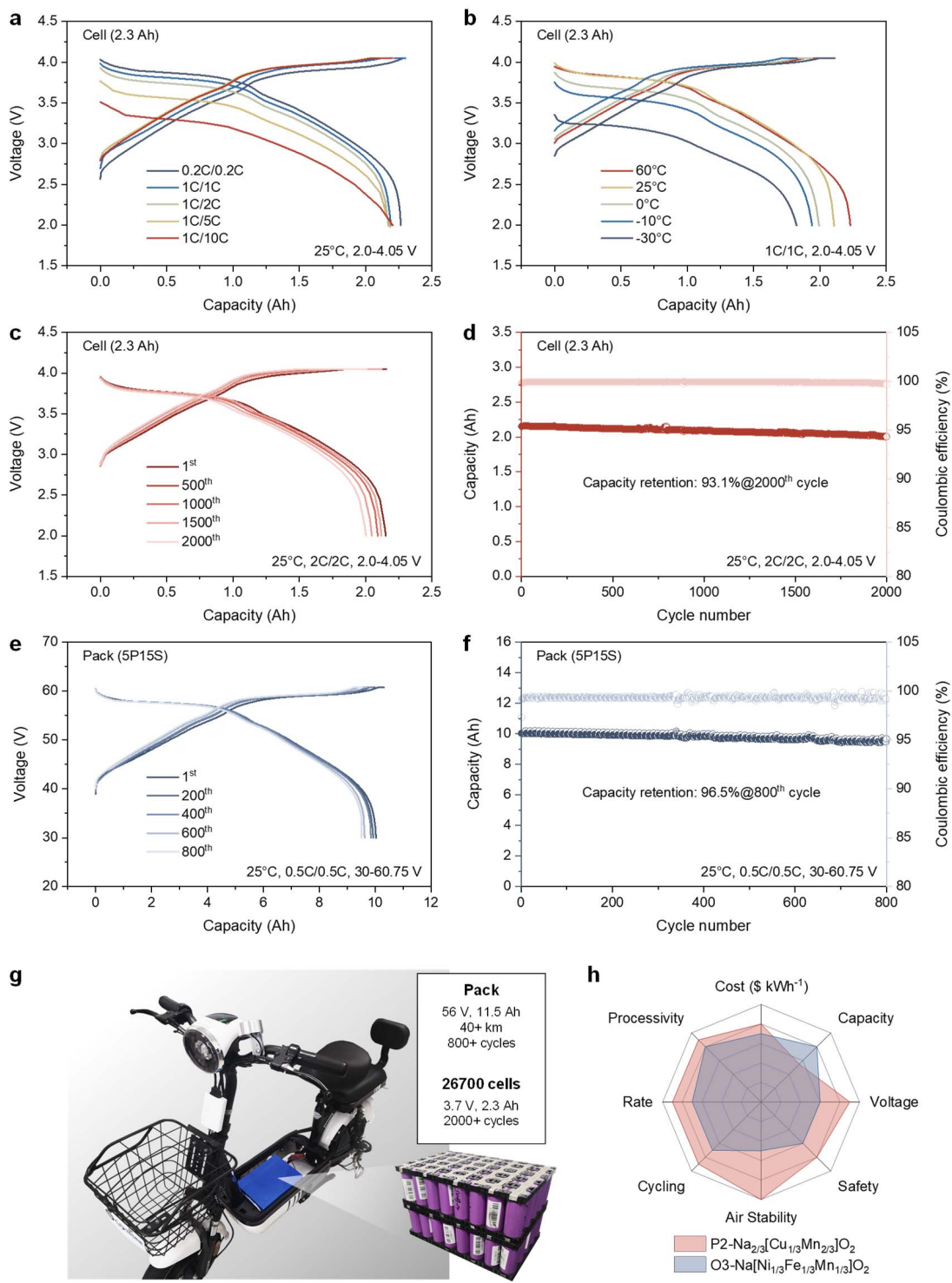

**Fig. 3 | Electrochemical performances of 26700 cylindrical cells and battery pack.**
**a**, Rate performance of 26700 cylindrical cells (P2-Na$_{2/3}$[Cu$_{1/3}$Mn$_{2/3}$]O$_2$ as the cathode material and hard carbon as the anode material), showing a high capacity retention of 97.2% and a high energy retention of 82.9% at a 10C rate (0.2C as 100%). **b**, Charge/discharge performance of 26700 cylindrical cells at various temperatures, showing a high capacity retention of 86.6% and a high energy retention of 73.4% at -30°C (25°C as 100%). **c**, **d**, Cycling performance of 26700 cylindrical cells. Charge/discharge voltage profiles (**c**) and plot of the discharge capacity and coulombic efficiency with the cycle number (**d**). **e**, **f**, Cycling performance of the battery pack of an E-bike. Charge/discharge voltage profiles (**e**) and plot of the discharge capacity and coulombic efficiency as a function of the cycle number (**f**). **g**, Test-related photographs of 26700 cylindrical cells, battery pack, and E-bike. **h**, Radar plot showing the performance characteristics of P2-$Na_{2/3}[Cu_{1/3}Mn_{2/3}]O_2$ and O3-$Na[Ni_{1/3}Fe_{1/3}Mn_{1/3}]O_2$.

**Table 1** Performance at various C-rates and temperatures of 26700 cylindrical cells

| No. | Charge rate/ Discharge rate | Temperature (°C) | Capacity retention (%) | Energy retention (%) | Discharge median voltage (V) |
|---|---|---|---|---|---|
| **Performance at various C-rates** | | | | | |
| 1 | 0.2C/0.2C | 25°C | 100.0% | 100.0% | 3.70 V |
| 2 | 1C/1C | 25°C | 96.6% | 95.7% | 3.66 V |
| 3 | 1C/2C | 25°C | 95.8% | 93.2% | 3.60 V |
| 4 | 1C/5C | 25°C | 96.4% | 89.1% | 3.43 V |
| 5 | 1C/10C | 25°C | 97.2% | 82.9% | 3.11 V |
| **Performance at various temperatures** | | | | | |
| 1 | 1C (25°C)/1C | 60°C | 105.8% | 104.0% | 3.59 V |
| 2 | 1C (25°C)/1C | 25°C | 100.0% | 100.0% | 3.66 V |
| 3 | 1C (25°C)/1C | 0°C | 94.6% | 91.5% | 3.55 V |
| 4 | 1C (25°C)/1C | -10°C | 92.1% | 86.4% | 3.43 V |
| 5 | 1C (25°C)/1C | -30°C | 86.6% | 73.4% | 3.08 V |

Note 1. All cells are of 26700 type;

Note 2. All cells are tested in the voltage range of 2.0-4.05 V after formation.

Note 3. For the cells tested at various temperatures, the cell is firstly charged at 25°C before discharging at various temperatures.